\documentclass[journal]{IEEEtran}

\usepackage{amsmath}
\usepackage{amssymb}
\usepackage{algorithm}
\usepackage{algpseudocode}
\usepackage{graphicx} 
\usepackage{color}
\usepackage{multirow}

\begin{document}
\pagestyle{plain}

\title{Strategies for CT Reconstruction using Diffusion Posterior Sampling with a Nonlinear Model}
\author{Xiao~Jiang,
        Shudong~Li,
        Peiqing~Teng,
        Grace~Gang,
        and~J.~Webster~Stayman

\thanks{This work has been submitted to the IEEE for possible publication. Copyright may be transferred without notice, after which this version may no longer be accessible.}
\thanks{Xiao Jiang and J.Webster Stayman are with the Department of Biomedical Engineering, Johns Hopkins University, Baltimore, MD, 20205 USA. e-mail: xjiang43@jhu.edu, web.stayman@jhu.edu}
\thanks{Shudong Li is with the Department of Electronic Engineering, Tsinghua University, Beijing, 100084 China. e-mail: sli248@jh.edu}
\thanks{Peiqing Teng is with the Department of Electrical and Computer Engineering, Johns Hopkins University, Baltimore, MD, 20205 USA. e-mail: pteng3@jhu.edu}
\thanks{Grace Gang is with the Department of Radiology, University of Pennsylvania, Philadelphia, PA, 19104 USA. e-mail: grace.j.gang@pennmedicine.upenn.edu}}

\maketitle

\begin{abstract}
Diffusion Posterior Sampling(DPS) methodology is a novel framework that permits nonlinear CT reconstruction by integrating a diffusion prior and an analytic physical system model, allowing for one-time training for different  applications. However, baseline DPS can struggle with large variability, hallucinations, and slow reconstruction. This work introduces a number of strategies designed to enhance the stability and efficiency of DPS CT reconstruction. Specifically, jumpstart sampling allows one to skip many reverse time steps, significantly reducing the reconstruction time as well as the sampling variability. Additionally, the likelihood update is modified to simplify the Jacobian computation and improve data consistency more efficiently. Finally, a hyperparameter sweep is conducted to investigate the effects of parameter tuning and to optimize the overall reconstruction performance. Simulation studies demonstrated that the proposed DPS technique achieves up to $46.72\%$ PSNR and $51.50\%$ SSIM enhancement in a low-mAs setting, and an over $31.43\%$ variability reduction in a sparse-view setting. Moreover, reconstruction time is sped up from $>23.5$ s/slice to $<1.5$ s/slice. In a physical data study, the proposed DPS exhibits robustness on an anthropomorphic phantom reconstruction which does not strictly follow the prior distribution. Quantitative analysis demonstrates that the proposed DPS can accommodate various dose levels and number of views. With $10\%$ dose, only a $5.60\%$ and $4.84\%$ reduction of PSNR and SSIM was observed for the proposed approach. Both simulation and phantom studies demonstrate that the proposed method can significantly improve reconstruction accuracy and reduce computational costs, greatly enhancing the practicality of DPS CT reconstruction.
\end{abstract}


\section{Introduction}

\IEEEPARstart{C}{omputed} Tomography (CT) reconstruction algorithms can generally be categorized into three types: analytical reconstruction\cite{feldkamp1984practical,noo2004two}, iterative reconstruction\cite{sidky2008image,yu2009compressed,chen2008prior}, and deep-learning-based reconstruction\cite{chen2017low,li2019learning,xia2023diffusion}. Analytical reconstruction derives explicit inversion formulas that directly map measurements to CT images, and have been widely used in clinical scanners\cite{pan2009commercial} due to the advantages of fast execution and low memory consumption. However, direct analytical methods require sufficiently high-quality measurements. Their performance can be significantly degraded by noise and biases not included in the idealized analytic model. To address this, such approaches often apply pre- or post-reconstruction filtration or other denoising/correction\cite{kyriakou2010empirical,zhu2009noise}. Iterative reconstruction addresses such issues by formulating the reconstruction as an optimization problem with accurate physics modeling and regularization to mitigate noise and encourage desirable image features. This kind of reconstruction can model many physical effects including noise to significantly enhance image quality in the case of low radiation exposures and irregular sampling \cite{stayman2013pirple,gang2022universal}. Nevertheless, most regularization strategies struggle to fully capture the complex distribution of human CT scans, and the image models based on such regularization can result in unrealistic image textures and features. Deep-learning reconstruction has gained increasing attention over the last decade. Large image databases can be used to train neural networks to capture the complexity of human CT scans. These neural networks can then be employed within the reconstruction process for tasks like projection-domain correction \cite{lee2018deep} and image-domain refinement\cite{tivnan2023fourier}, or for direct mapping from projections to images\cite{chen2018learn,he2020radon}, offering fast processing and improved image quality over traditional approaches. 

In clinical use, CT scan protocols, can vary significantly with patient size/habitus, imaging task, and anatomical target. Despite the great success of deep learning methods, most current networks are trained on datasets tailored to a specific system configuration. This potentially results in reduced performance when applied to a new protocol or necessitates model retraining to accommodate system changes. To address this issue, recent studies have explored training a single network with different imaging conditions to improves the model generalization ability\cite{maier2019real}. Xia \textit{et al.} \cite{xia2021ct} and Wang \textit{et al.}\cite{wang2021ct} improve this kind of training by introducing auxiliary variables to modulate network processing, and adapting to a range of imaging conditions. The advent of score-based generative models (SGMs) has offered an alternative strategy\cite{xia2023diffusion,song2020score}, which uses a generative network to capture a target-domain/clean-image distribution, then to combine that generative model with an analytic physical model of the scanner to reconstruct CT images. In effect, these approaches combine the flexible physical modeling of iterative reconstruction with the sophisticated image information provided by deep learned priors. A key advantage of this approach is its independence from paired training data and specific imaging conditions, which significantly simplifies the network training process and elevates the ability to generalize to new applications.

While SGM-based reconstruction shows promising performance, most existing research focuses on a linear forward model\cite{chung2023fast,xia2023diffusion,chung2023solving}. However, accurately representing the intricacies of CT imaging physics—such as system blur\cite{tilley2017penalized}, spectral modeling\cite{tilley2019model}, and scattered radiation effects \cite{zhu2006scatter} —requires a shift towards nonlinear modeling. Recent work on Diffusion Posterior Sampling (DPS) \cite{chung2022diffusion} provides a general framework to solve nonlinear inverse problem with a SGM-based algorithm. Our investigations into DPS CT reconstruction\cite{li2023diffusion}, incorporating a nonlinear physics model, have shown encouraging results in simulation studies. Despite these advancements, the many computations associated with the reverse sampling steps and the inherent stochasticity of the DPS approach still raises concerns regarding the speed and stability of reconstructions.

In this work, we consider the drawbacks of the originally proposed DPS nonlinear CT reconstruction and investigate improved strategies to shorten the reconstruction time as well as to reduce sampling variability. Both simulation and physical phantom studies are conducted to evaluate the proposed strategies. The paper is organized as follows: Sec.\ref{sec:DPS} introduces the framework of DPS nonlinear CT reconstruction. Sec.\ref{sec:JS} and Sec.\ref{sec:ML} introduce the improved strategies for stabilized and accelerated DPS CT reconstruction. Sec.\ref{sec:DDPMtrain} and Sec.\ref{sec:Eval} present implementation details including model training, experiment setup, and evaluation metrics. Experimental results and analysis are included in Sec.\ref{sec:Res}, with conclusion and discussion summarized in Sec.\ref{sec:Dis}. Some of the strategies introduced in this work were briefly reported in conference proceedings \cite{jiang2024ct} for a related DPS approach for spectral CT. This work includes additional strategies and evaluations focusing on single-energy CT reconstruction. 

\section{Methodology}
\subsection{Diffusion Posterior Sampling CT Reconstruction}
\label{sec:DPS}
\subsubsection{Nonlinear CT Forward Model}
In this work, we use the general nonlinear CT forward model proposed in Ref. \cite{tilley2017penalized}:
\begin{subequations}
\label{eq:phymodel}
\begin{equation}
\label{eq:noisymodel}
    \textbf{y} \sim \mathcal{N}(\overline{\textbf{y}}, \textbf{K})
\end{equation}
\begin{equation}
\label{eq:meanmodel}
    \overline{\textbf{y}}=\textbf{B}\exp\{-\textbf{Ax}\}.
\end{equation}
\end{subequations}

\noindent The measurements $\textbf{y}$ are assumed to follow a multivariate Gaussian distribution with mean $\overline{\textbf{y}}$ and covariance $\textbf{K}$. In the mean measurement model Eq.\eqref{eq:meanmodel}, $\textbf{x}$ represents the attenuation map to be reconstructed, and the projection matrix $\textbf{A}$ characterizes the system geometry. The matrix $\textbf{B}$ encapsulates both the pixel-dependent photon fluence and gain factors, as well as the system blur, which we consider to be shift-invariant in this work. The Gaussian distribution of $\textbf{y}$ results in the following log-likelihood objective function for estimation of $\textbf{x}$:
\begin{equation}
\label{eq:ML}
    -\log p(\textbf{y}|\textbf{x}) \propto \|\textbf{B}\exp(-\textbf{Ax})-\textbf{y}\|_{\textbf{K}^{-1}}^2.
\end{equation}

\subsubsection{Diffusion Posterior Sampling with a Nonlinear Forward Model}
The Score-based Generative Model (SGM)\cite{song2020score} provides an novel approach for generating new samples from a target domain. It is founded on the principle of a forward process that incrementally applies time-dependent noise to a sample from the target domain and a reverse process that gradually removes the image noise as estimated by neural network. This bidirectional methodology is rigorously defined through paired stochastic differential equations (SDEs). Specifically, the SDEs for Denoising Diffusion Probabilistic Model (DDPM) \cite{ho2020denoising} have the following form:

\begin{subequations}
\begin{equation}
\label{eq:forward}
    \text{Forward}: \mathrm{d}\textbf{x} = -\frac{\beta_t}{2} \textbf{x}\mathrm{d}t + \sqrt{\beta_t}\mathrm{d}\textbf{w}  
\end{equation}
\begin{equation}
\label{eq:reverse}
    \text{Reverse}: \mathrm{d}\textbf{x} = [-\frac{\beta_t}{2} \textbf{x}-\beta_t\nabla_{\textbf{x}_t} \mathrm{log} p_t({\textbf{x}_t})]\mathrm{d}t + \sqrt{\beta_t}\mathrm{d}\textbf{w}.
\end{equation}
\end{subequations}

\noindent where $\beta_t$ is the variance scheduler of the incremental noise and $\mathrm{d}\textbf{w}$ is the standard Wiener process. A deep neural network is used to approximate the unknown score function $\textbf{s}_\theta(\textbf{x}_t,t) \approx \nabla_{\textbf{x}_t} \mathrm{log}p_t({\textbf{x}_t})$. The reverse process can be directly generalized to conditional generation as follows:

\begin{equation}
\label{eq:condition}
    \mathrm{d}\textbf{x} = [-\frac{\beta_t}{2} \textbf{x}-\beta_t\nabla_{\textbf{x}_t} \mathrm{log} p_t({\textbf{x}_t|\textbf{y}})]\mathrm{d}t + \sqrt{\beta_t}\mathrm{d}\textbf{w}.
\end{equation}

In the context of CT imaging reconstruction, $\textbf{y}$ and $\textbf{x}$ denote the measurements and the reconstructed images as defined in Eqs.\eqref{eq:noisymodel},\eqref{eq:meanmodel}. Diffusion Posterior Sampling (DPS) \cite{chung2022diffusion} reformulates the conditional sampling by leveraging Bayes rule $p(\textbf{x}|\textbf{y}) \propto p(\textbf{x})p(\textbf{y}|\textbf{x})$: 

\begin{equation}
\label{eq:bayes}
\begin{aligned}
    \mathrm{d}\textbf{x} = &[-\frac{\beta_t}{2} \textbf{x}-\beta_t\nabla_{\textbf{x}_t} \mathrm{log} p_t({\textbf{x}_t})]\mathrm{d}t + \sqrt{\beta_t}\mathrm{d}\textbf{w} \\
    &- \beta_t\nabla_{\textbf{x}_t} \mathrm{log} p_t({\textbf{y}|\textbf{x}_t})\mathrm{d}t.
\end{aligned}
\end{equation}

\noindent The conditional distribution $p_t({\textbf{y}|\textbf{x}_t})$, which is generally intractable for the nonlinear problem, can be approximated based on Tweedie’s formula \cite{chung2022diffusion,efron2011tweedie}:
\begin{equation}
\label{eq:dps_appro}
    p_t(\textbf{y}|\textbf{x}_t) \approx p_t(\textbf{y}|\hat{\textbf{x}}_0), \text{where} \ \hat{\textbf{x}}_0 = 
\frac{1}{\sqrt{\overline{\alpha}_t}}(\textbf{x}_t+(1-\overline{\alpha}_t))\textbf{s}_\theta(\textbf{x}_t,t).
\end{equation}

\noindent Combining the approximation \eqref{eq:dps_appro} and likelihood function \eqref{eq:ML}, we have DPS for nonlinear CT reconstruction \cite{li2023diffusion}:
\begin{equation}
\label{eq:dps_ct}
\begin{aligned}
    \mathrm{d}\textbf{x} = &[-\frac{\beta_t}{2} \textbf{x}-\beta_t\nabla_{\textbf{x}_t} \mathrm{log} p_t({\textbf{x}_t})]\mathrm{d}t + \sqrt{\beta_t}\mathrm{d}\textbf{w} \\
    &+ \beta_t \nabla_{\textbf{x}_t} {\hat{\textbf{x}}_0} \nabla_{\hat{\textbf{x}}_0} \|\textbf{B}\exp\{-\textbf{A}\hat{\textbf{x}}_0\}-\textbf{y}\|_{\textbf{K}^{-1}}^2 \mathrm{d}t.
\end{aligned}
\end{equation}
\noindent The top part of Eq.\eqref{eq:dps_ct} is exactly same as unconditional diffusion sampling, which introduces the prior information captured by the diffusion model, and the bottom part incorporates the physics model to enhance the data consistency. It is worth noting that modifications to the physics model impact only the likelihood term, making it feasible to employ a single unconditional DDPM model, trained solely on high-quality CT images, to do image reconstruction for different CT imaging conditions, e.g., different physical forward models, without additional training.

\subsubsection{Discretized Implementation}
In practice, The continuous forward process is discretized into $T$ time steps \cite{ho2020denoising}:
\begin{equation}
\label{eq:ddpm_f}
\begin{aligned}
    &\textbf{x}_t = \sqrt{\bar{\alpha}_t}\textbf{x}_0 + \sqrt{1-\bar{\alpha}_t}\boldsymbol{\epsilon} \\
    \text{where} &\ \bar{\alpha}_t=\prod_{i=1}^{t} (1-\beta_t), \ \boldsymbol{\epsilon} \sim \mathcal{N}(0,\boldsymbol{I}), \ t=1,2...T.
\end{aligned}
\end{equation}

\noindent Since the noise $\boldsymbol{\epsilon}$ is related to the score function by $\boldsymbol{\epsilon} = -\sqrt{1-\bar{\alpha}_t} \ \mathrm{log} p_t({\textbf{x}_t})$, DDPM actually trains a network $\boldsymbol{\epsilon}_\theta(\textbf{x},t)$ to predict the noise $\boldsymbol{\epsilon}$:
\begin{equation}
\label{eq:ddpm_train}
    \theta^* =  \text{argmin} \ \mathbb{E}_{\textbf{x}_0}\mathbb{E}_{\boldsymbol{\epsilon},t} \|\boldsymbol{\epsilon}_\theta(\textbf{x}_t,t)-\boldsymbol{\epsilon}\|_2^2.
\end{equation}
Using the trained network and physics model, the image $\textbf{x}_0$ may be reconstructed by solving the reverse SDE \cite{ho2020denoising,song2020denoising}, as outlined in Algorithm 1. Consistent with the approach adopted in Ref.\cite{chung2022diffusion}, we utilize an adjustable parameter $\eta_t$ instead of variance scheduler $\beta_t$ to control the step size of gradient descent.

\begin{algorithm}
\caption{Baseline DPS}
\begin{algorithmic}[1]\small
\State $T$: diffusion steps
\State $\eta_t$: step size
\State
\State $\textbf{x}_T\sim\mathcal{N}~(0,\boldsymbol{I})$
\For{\texttt{$t = T$ to $1$}}:
    \State $\textbf{z} \sim \mathcal{N}~(0,\boldsymbol{I})$
    \State
    \State \# Diffusion sampling:
    \State $\hat{\textbf{x}}_0 = \frac{1}{\sqrt{\bar{\alpha}_t}}(\textbf{x}_t-\sqrt{1-\bar{\alpha}_t}\boldsymbol{\epsilon}_\theta(\textbf{x}_t,t))$
    \State $\textbf{x}_{t-1}' = \frac{\sqrt{\alpha_t}(1 - \bar{\alpha}_{t-1})}{1 - \bar{\alpha}_t} \textbf{x}_t + 
    \frac{\sqrt{\bar{\alpha}_{t-1}}\beta_t} {1 - \bar{\alpha}_t} \hat{\textbf{x}}_0 + \sigma_t \textbf{z}$
    \State
    \State \# Likelihood update:
    \State $\textbf{x}_{t-1} = \textbf{x}_{t-1}' - \eta_t \nabla_{\textbf{x}_t} {\hat{\textbf{x}}_0}\nabla_{\hat{\textbf{x}}_0} \left\| \textbf{B}\exp({-\textbf{A}\hat{\textbf{x}}_0}) -\textbf{y}\right\|_{\textbf{K}^{-1}}^2 $ 
\EndFor
\end{algorithmic}
\end{algorithm}

\begin{figure*}[h]
\centering
\includegraphics[width=\textwidth]{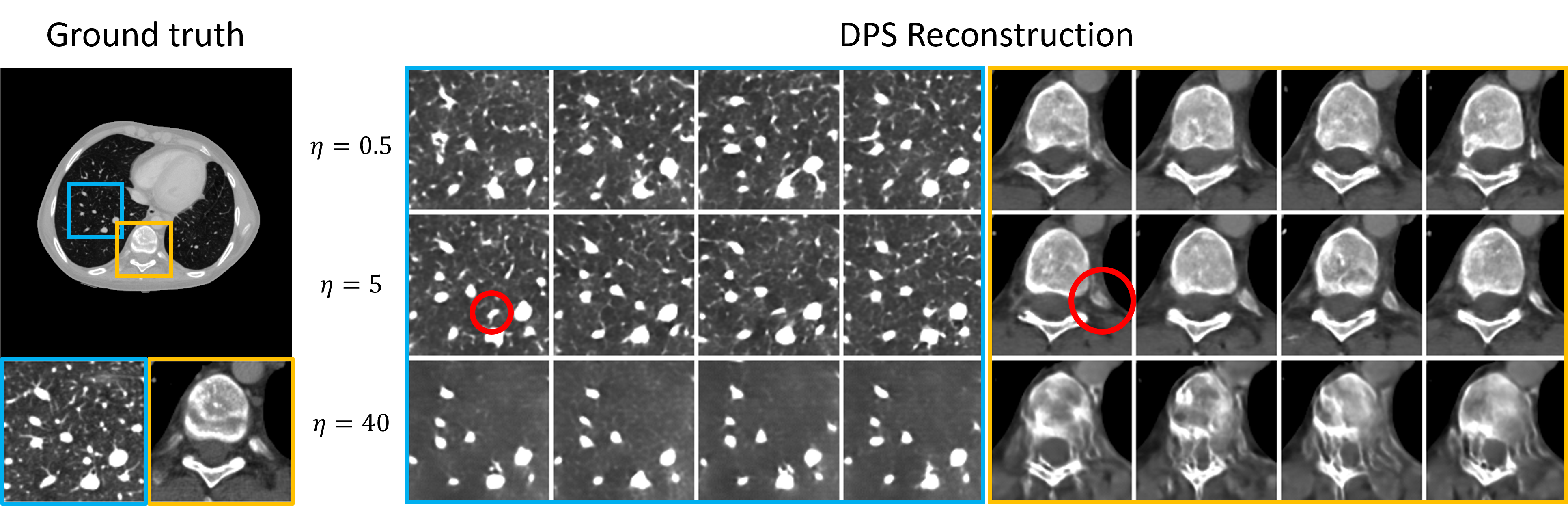}
\caption{Effects of step size on DPS reconstruction for low-dose ($I_0=5000$) measurements. The step size is determined as $\eta_t = \eta/\|\nabla_{\textbf{x}_t} {\hat{\textbf{x}}_0}\nabla_{\hat{\textbf{x}}_0} \left\| \textbf{B}\exp({-\textbf{A}\hat{\textbf{x}}_0}) -\textbf{y}\right\|_{\textbf{K}^{-1}}^2\|$ \cite{li2024ctreconstructionusingdiffusion}. For each step size, DPS reconstructions are repeated four times for the same measurements, and the results of lung ROI and spine ROI are zoomed in for variability analysis. Large $\eta=40$ places more emphasis on data consistency, resulting in unrealistic images, while small $\eta=0.5$ tends to introduce considerable structural bias and variability. Optimized $\eta=5$ can stabilize the reconstruction but still suffers substantial variability in some details.}
\label{fig:dps} 
\end{figure*}

Algorithm 1 provides a general reconstruction framework to integrate the sophisticated diffusion prior and accurate CT physics model. However, it still faces a few practical challenges:
\begin{itemize}
  \item DDPM can require thousands of steps to generate high-quality samples from pure noise, resulting in a time-consuming CT reconstruction process. Similarly, computation of the Jacobian $\nabla_{\textbf{x}_t} {\hat{\textbf{x}}_0}$ also brings significant memory and computation time requirements due to the gradient backpropagation through the neural network.
  \item Previous studies \cite{chung2022diffusion,li2023diffusion} have highlighted the importance of carefully scheduling the step size $\eta_t$. A large $\eta_t$ (Fig.\ref{fig:dps}, bottom row) can interfere with stable evolution of the diffusion distribution, while a small $\eta_t$ (Fig.\ref{fig:dps}, top row) may lead to inadequate incorporation of measurement information, making the reconstruction prone to generate hallucinations.
  \item Due to the inherent stochasticity of diffusion sampling, reconstruction results can exhibit substantial variability even with optimized step size (Fig.\ref{fig:dps}, middle row), This inconsistency presents a significant hurdle for applying DPS in clinical CT reconstruction.
\end{itemize}

\noindent In the following sections, we will introduce several strategies for stabilized and accelerated DPS CT reconstruction.

\subsection{Jumpstart Strategy}
\label{sec:JS}
During DPS reconstruction, diffusion sampling introduces stochastic perturbations at each time step, providing variation in the sampling outcomes, while the likelihood update enhances data consistency, thereby "refocusing" the sampling trajectory. Step size acts as a relative weighting between prior information and data consistency. However, as illustrated in Fig.\ref{fig:dps}, it can be challenging to select an appropriate step size that ensures both stability and accuracy. Since the reconstruction variability is associated with the diffusion sampling, variation can be greatly reduced if sampling is started from an intermediate state $\textbf{x}_{T'}$ which avoids divergence over hundreds of starting steps. In order to obtain an appropriate $\textbf{x}_{T'}$ that follows the prior distribution $p(\textbf{x}_{T'})$ but also remains consistent with the measurements, one may start reconstruction process with an initial estimate $\hat{\textbf{x}}_0^f$ which is close to the solution $\textbf{x}_0$. Applying forward diffusion to $\textbf{x}_0$ and $\hat{\textbf{x}}_0^f$ results in the following distribution:
\begin{subequations}
\begin{equation}
    p(\textbf{x}_t|\textbf{x}_0) \sim \mathcal{N}(\sqrt{\bar{\alpha}}_t\textbf{x}_0,(1-\bar{\alpha}_t)\textbf{I})
\end{equation}
\begin{equation}
    p(\hat{\textbf{x}}_t^f|\hat{\textbf{x}}_0^f) \sim \mathcal{N}(\sqrt{\bar{\alpha}}_t\hat{\textbf{x}}_0^f,(1-\bar{\alpha}_t)\textbf{I}).
\end{equation}
\end{subequations}

\noindent The discrepancy between $\textbf{x}_t$ and $\hat{\textbf{x}}_t^f$ can be quantified by the KL divergence:
\begin{equation}
\label{eq:KL}
D_{KL}(p(\textbf{x}_t|\textbf{x}_0)\parallel p(\hat{\textbf{x}}_t^f|\hat{\textbf{x}}_0^f))=\frac{\bar{\alpha}_t}{2(1-\bar{\alpha}_t)}\|\hat{\textbf{x}}_0^f-\textbf{x}_0\|_2^2.
\end{equation}

\noindent Although $\hat{\textbf{x}}_0^f$ and $\textbf{x}_0$ might follow different distributions, Eq.\eqref{eq:KL} demonstrates that this discrepancy diminishes as the forward diffusion progresses because ${\bar{\alpha}_t}/{(1-\bar{\alpha}_t)}$ progressively shrinks towards zero. For sufficiently large $t$, it is reasonable to assume $\textbf{x}_t$ and $\hat{\textbf{x}}_t^f$ conform to nearly identical distributions. This convergence suggests that $\hat{\textbf{x}}_t^f$ can serve as intermediate state image and be used to initiate reverse sampling. 

\begin{figure}[h]
\centering
\includegraphics[width=\linewidth]{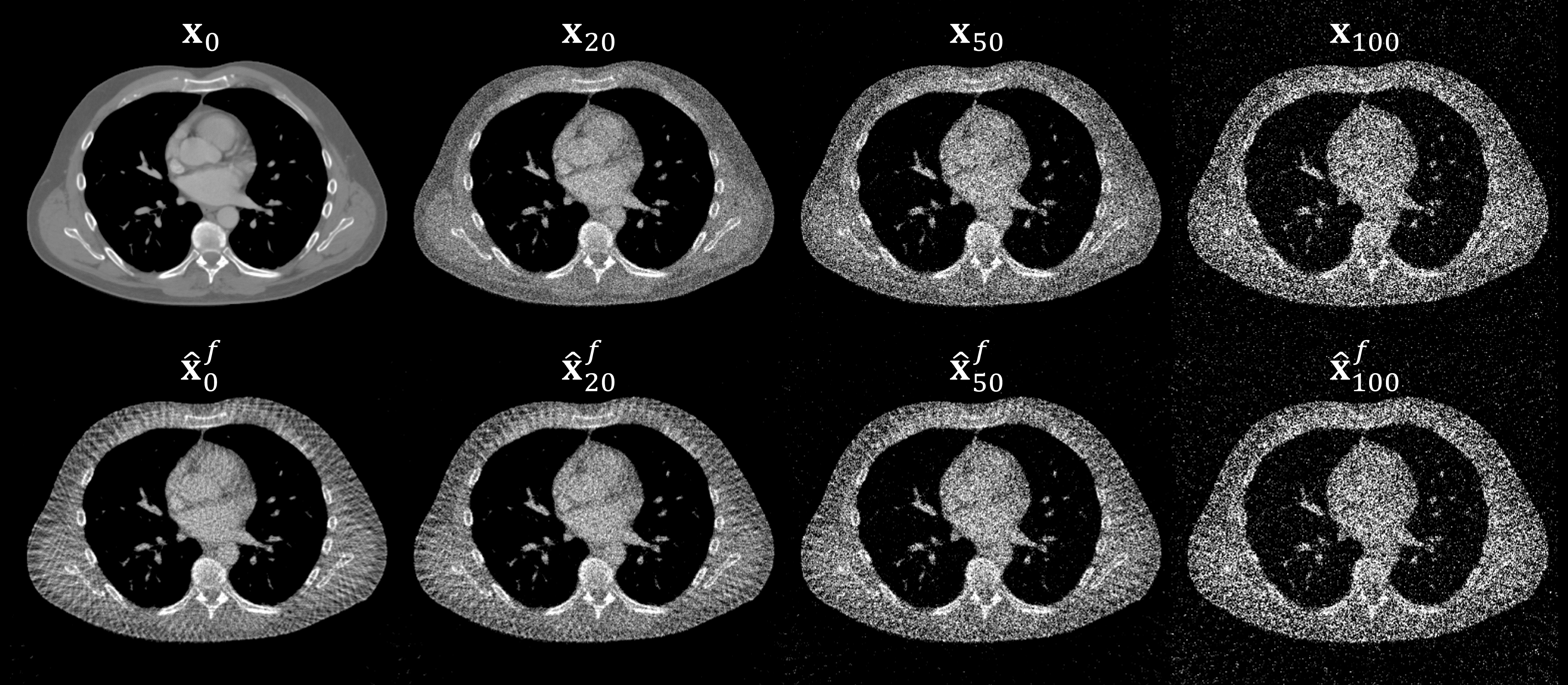}
\caption{Forward diffusion process starting from the ground truth (top) and a noisy FBP reconstruction on sparse (72 view) projections (bottom), respectively. The left column images shows a significant difference. However, as the diffusion progresses, $\textbf{x}_t$ and $\hat{\textbf{x}}_t^f$ tend to follow similar distributions. The forward diffusion in this example uses a total of 1000 steps, while $\sim$ 100 steps can effectively mitigate the discrepancy between FBP reconstruction and the ground truth.}
\label{fig:js} 
\end{figure}

Chung et al. \cite{chung2022come} proposed similar idea to solve linear problem, and trained a neural network to provide the initial estimate. Here we extend this idea to the nonlinear problem, and apply Filtered-backprojection (FBP) reconstruction as an initial estimate. Such FBP images are generally available through fast computation on current CT systems. Fig.\ref{fig:js} illustrates that even when FBP results are severely corrupted by noise and streaking artifacts, a sufficient number of time step (e.g., $T'\ll T$) can effectively mitigate the discrepancy between $\textbf{x}_t$ and $\hat{\textbf{x}}_t^f$. Consequently, this approach allows ``jumpstart" over many early time steps, which not only reduces the sampling divergence, but also significantly speeds up the reconstruction.

\begin{figure*}[ht]
\centering
\includegraphics[width=\textwidth]{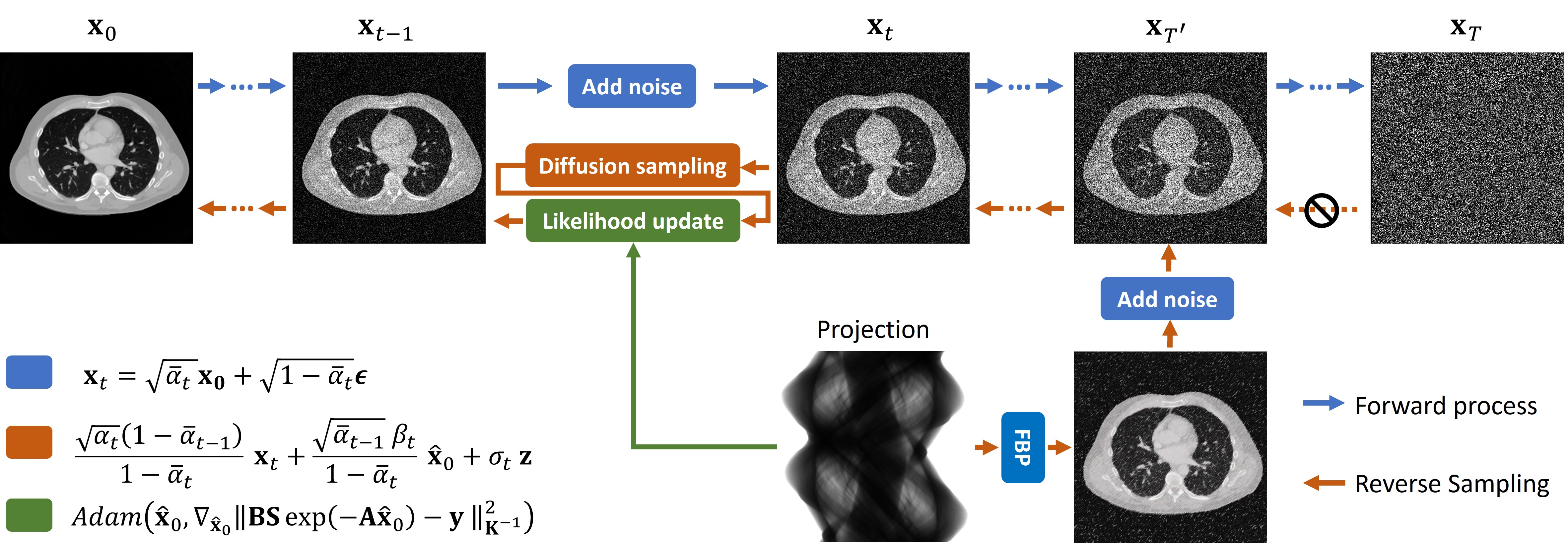}
\caption{Algorithm 2 Workflow: The blur arrows and brown arrows represent the forward diffusion process for network training and the proposed reverse sampling for CT reconstruction, respectively. The dotted arrow indicates the initial hundreds of reverse diffusion steps bypassed by the jumpstart strategy.}
\label{fig:workflow} 
\end{figure*}

\subsection{Modified Likelihood Update}
\label{sec:ML}
\subsubsection{Jacobian Approximation}
The likelihood update in Algorithm 1 involves the computation of posterior-mean Jacobian and likelihood gradient:
\begin{equation}
\label{eq:jacobdecomp}
\underbrace{\nabla_{\textbf{x}_t} {\hat{\textbf{x}}_0}}_\textit{Posterior-mean Jacobian} \underbrace{\nabla_{\hat{\textbf{x}}_0} \left\| \textbf{B}\exp({-\textbf{A}\hat{\textbf{x}}_0}) -\textbf{y}\right\|_{\textbf{K}^{-1}}^2}_\textit{Likelihood gradient}
\end{equation}

\noindent The posterior-mean Jacobian can be expanded as:
\begin{equation}
\label{eq:jacob}
\nabla_{\textbf{x}_t}{\hat{\textbf{x}}_0} = \frac{1}{\sqrt{\bar{\alpha}_t}}(\mathbf{I}-\sqrt{1-\bar{\alpha}_t}\nabla_{\textbf{x}_t}\boldsymbol{\epsilon}_\theta(\textbf{x}_t,t)).
\end{equation}

\noindent Computation of Eq.\eqref{eq:jacob} requires gradient backpropagation through the deep neural network $\boldsymbol{\epsilon}_\theta$, which is both time- and memory-intensive. However, based on numerical evaluation (detailed in Appendix \ref{sec:append_jacob}), we find the network Jacobian $\nabla_{\textbf{x}_t}\boldsymbol{\epsilon}_\theta(\textbf{x}_t,t))$ can be well approximated by a diagonal matrix $\frac{\textbf{I}}{\sqrt{1-\bar{\alpha}_t}}$. This indicates the posterior mean Jacobian $\nabla_{\textbf{x}_t}{\hat{\textbf{x}}_0}$ can be poorly-conditioned and its computation is often unstable as suggested in previous studies\cite{poole2022dreamfusion,du2023reduce}. To address these issues, we propose to discard the second term in the Eq.\eqref{eq:jacob}, bypassing the network Jacobian calculation with a stable estimate:
\begin{equation}
\label{eq:jacobappro}
\nabla_{\textbf{x}_t}{\hat{\textbf{x}}_0} \rightarrow \frac{\textbf{I}}{\sqrt{\bar{\alpha}_t}}.
\end{equation}
\noindent The overall likelihood update then can be simplified as:
\begin{align}
\label{eq:gardappro}
\textbf{x}_{t-1} & = \textbf{x}_{t-1}'  - \eta_t \frac{\textbf{I}}{\sqrt{\bar{\alpha}_t}} \nabla_{\hat{\textbf{x}}_0} \left\| \textbf{B}\exp({-\textbf{A}\hat{\textbf{x}}_0}) - \textbf{y}\right\|_{\textbf{K}^{-1}}^2
\notag \\
                 & = \textbf{x}_{t-1}'  - \eta_t \nabla_{\hat{\textbf{x}}_0} \left\| \textbf{B}\exp({-\textbf{A}\hat{\textbf{x}}_0}) - \textbf{y}\right\|_{\textbf{K}^{-1}}^2.
\end{align}

\noindent The second row of Eq.\eqref{eq:gardappro} further absorbs Jacobian scalar factors into the user-selectable step size. 

\subsubsection{Multi-Step Likelihood Update}
Leveraging Jacobian approximation, the likelihood update Eq.\eqref{eq:gardappro} essentially performs a descent along the likelihood gradient on $\hat{\textbf{x}}_0$. We slightly reformulate Eq.\eqref{eq:gardappro}:
\begin{align}
\label{eq:reform}
    \textbf{x}_{t-1} & = \textbf{x}_{t-1}'  - \hat{\textbf{x}}_0 + \hat{\textbf{x}}_0 - \eta_t \nabla_{\hat{\textbf{x}}_0} \left\| \textbf{B}\exp({-\textbf{A}\hat{\textbf{x}}_0}) - \textbf{y}\right\|_{\textbf{K}^{-1}}^2 \notag \\
                     & = \textbf{x}_{t-1}'  - \hat{\textbf{x}}_0 + \hat{\textbf{x}}_0',
\end{align}
where $\hat{\textbf{x}}_0'$ can be considered as an updated $\hat{\textbf{x}}_0$ with enhanced data consistency. For the ill-conditioned nonlinear CT reconstruction problem, achieving adequate data consistency through a single-step gradient descent per time step can prove challenging, especially for DPS reconstructions which perturb the image with independent noise in each time step. To facilitate stable and substantial consistency enhancement, we apply a multi-step gradient descent per time step. In order to effectively manage the step size and accelerate the convergence, we borrow from large-scale optimization, implementing gradient descent with Adaptive Moment Estimation (Adam) algorithm \cite{kingma2014adam}, which facilitates a smooth optimization trajectory and allows for adaptive step size adjustment. Experimental results demonstrate that $2\sim5$ Adam updates per time step are adequate to achieve satisfactory consistency. The increased computational demand associated with a multi-step update can be further offset by leveraging the ordered subset strategy\cite{erdogan1999ordered}. Specifically, it is well-known in model-based tomographic reconstruction that an appropriate gradient may be computed using a fraction of the projections, i.e.:
\begin{equation}
\label{eq:os}
    \nabla_{\textbf{x}_0}\mathcal{L} (\textbf{x}_0,\textbf{y}) \approx S\nabla_{\textbf{x}_0}\mathcal{L} (\textbf{x}_0,\textbf{y}_s),
\end{equation}
where $S$ and $s$ is the number of subsets and the subset index, respectively. 

Integrating the jumpstart strategy and the modified likelihood updates, we propose a stable DPS for nonlinear CT reconstruction as outlined in Algorithm 2 and Fig. \ref{fig:workflow}.

\begin{algorithm}
\caption{Proposed Stable DPS}
\begin{algorithmic}[1]\small
\State $T$: diffusion training steps
\State $T'$: jumpstart steps, $T'\ll T$
\State $\hat{\textbf{x}}_0^f$: FBP reconstruction
\State $S$: number of subsets/number of likelihood update per time step
\State
\State \# Adam optimizer 
\State $\eta$: step size
\State $\gamma_1 = 0.9, \gamma_2 = 0.999$: momentum coefficients 
\State
\State \# Order subset
\State $\{\textbf{A}_s, \textbf{y}_s\}$: subset forward projector and measurements
\State
\State \# Reverse sampling initialization
\State $\boldsymbol{\epsilon}\sim\mathcal{N}~(0,\boldsymbol{I})$
\State $\textbf{x}_{T'} = \hat{\textbf{x}}_0^{T'}=\sqrt{\bar{\alpha}_{T'}}\hat{\textbf{x}}_0^f + \sqrt{1-\bar{\alpha}_{T'}}\boldsymbol{\epsilon}$
\State
\State \# Diffusion Posterior Sampling
\For{\texttt{$t = T'$ to $1$}}:
    \State \# Diffusion sampling:
    \State $\textbf{z} \sim \mathcal{N}~(0,\boldsymbol{I})$
    \State $\hat{\textbf{x}}_0 = \frac{1}{\sqrt{\bar{\alpha}_t}}(\textbf{x}_t-\sqrt{1-\bar{\alpha}_t}\boldsymbol{\epsilon}_\theta(\textbf{x}_t,t))$
    \State $\textbf{x}_{t-1}' = \frac{\sqrt{\alpha_t}(1 - \bar{\alpha}_{t-1})}{1 - \bar{\alpha}_t} \textbf{x}_t + 
    \frac{\sqrt{\bar{\alpha}_{t-1}}\beta_t} {1 - \bar{\alpha}_t} \hat{\textbf{x}}_0 + \sigma_t \textbf{z}$
    \State
    \State \# Likelihood update:
    \State $\hat{\textbf{x}}_0' = \hat{\textbf{x}}_0$ 
    \For{\texttt{$s = 1$ to $S$}}:
        \State $\hat{\textbf{x}}_0' = \text{Adam}(\hat{\textbf{x}}_0',S\nabla_{\hat{\textbf{x}}_0'}\left\| \textbf{B}\exp({-\textbf{A}_s\hat{\textbf{x}}_0'}) -\textbf{y}_s\right\|_{\textbf{K}^{-1}}^2, \gamma_1, \gamma_2)$
    \EndFor
    \State $\textbf{x}_{t-1} = \textbf{x}_{t-1}' - \hat{\textbf{x}}_0 + \hat{\textbf{x}}_0'$
\EndFor
\end{algorithmic}
\end{algorithm}

\subsection{Implementation Details}
\label{sec:DDPMtrain}
\subsubsection{Unconditional DDPM Training}
DPS reconstruction requires unconditional score network training. The training dataset is constructed based on the public CT Lymph Nodes dataset\cite{roth2014new}. The dataset contains torso CT scans of 150 patients, from which we extract the slices between neck and liver, forming a chest CT dataset of 30000 slices, each slice has $512\times512$ pixels. Those images are pre-processed to convert Hounsfield Units to attenuation coefficients and remove the patient table. 25000 slices of the processed images are used for model training, and the slices from patients excluded in the training dataset are used for reconstruction evaluation.

In this work, DDPM\cite{ho2020denoising} with a Residual Unet\cite{jiang2022residual} backbone is employed as the SGM framework. The continuous diffusion process is discretized into $T = 1000$ time steps with a linear variance scheduler from $\beta_1 = 1e^{-4}$ to $\beta_{1000} = 0.02$. The neural network was implemented using the PyTorch framework, and the Adam optimizer was used to minimize the loss function defined in Eq.\eqref{eq:ddpm_train} with a batch size of $16$ and a learning rate of $10^{-4}$. The training was terminated after 200 epochs.

\subsubsection{Parameter Selection}
The proposed algorithm incorporates three hyperparameters, the jumpstart steps $T'$, the likelihood update step size $\eta$, and the number of likelihood update per time step $S$. We conducted a comprehensive parameter sweep for $T'$, $\eta$, and $S$. This allowed us to explore the effects of these parameters on the imaging accuracy and reconstruction variability, and to illustrate how to fine tune the hyperparameters for accurate and stable reconstructions. The baseline DPS was implemented as described in Algorithm 1 for comparison. The step-size scheduler was designed as suggested in [33]: $\eta_t = \eta/\|\nabla_{\textbf{x}_t} {\hat{\textbf{x}}_0}\nabla_{\hat{\textbf{x}}_0} \left\| \textbf{B}\exp({-\textbf{A}\hat{\textbf{x}}_0}) -\textbf{y}\right\|_{\textbf{K}^{-1}}^2\|$. The constant $\eta$ is optimized to minimize the reconstruction variability.

\subsection{Evaluation}
\label{sec:Eval}
\subsubsection{Quantitative Metrics}
The overall reconstruction quality was quantified by peak signal to noise ratio (PSNR), and structural similarity index measure (SSIM) \cite{sara2019image}:
\begin{subequations}  
\begin{equation}
\label{eq:PSNR}
\text{PSNR}(\hat{\textbf{x}},\textbf{x}) = 10 \cdot \log_{10} \left( \frac{MAX_\textbf{x}^2}{MSE(\hat{\textbf{x}},\textbf{x})} \right)
\end{equation}
\begin{equation}
\label{eq:SSIM}
\text{SSIM}(\hat{\textbf{x}},\textbf{x}) = \frac{(2\mu_{\hat{\textbf{x}}}\mu_\textbf{x} + c_1)(2\sigma_{\hat{\textbf{x}}\textbf{x}} + c_2)}{(\mu_{\hat{\textbf{x}}}^2 + \mu_\textbf{x}^2 + c_1)(\sigma_{\hat{\textbf{x}}}^2 + \sigma_\textbf{x}^2 + c_2)},
\end{equation}
\end{subequations}

\noindent where $\hat{\textbf{x}}, \textbf{x}$ represent the reconstruction and ground truth image, respectively. Because of the inherent stochasticity, diffusion-based algorithms produce different images even when conditioned on the same measurements. Here we investigate such variability by computing bias and standard deviation (STD) maps on 32 independent DPS reconstructions using the same measurements:
\begin{subequations}  
\begin{equation}
\label{eq:Bias}
\text{Bias}(\hat{\textbf{x}},\textbf{x}) = |\mathbb{E}\{\hat{\textbf{x}}\} - \textbf{x}|
\end{equation}
\begin{equation}
\label{eq:STD}
\text{STD}(\hat{\textbf{x}}) = \sqrt{\mathbb{E}\{(\hat{\textbf{x}} - \mathbb{E}\{\hat{\textbf{x}}\})^2\}}
\end{equation}
\end{subequations}
\noindent An overall bias and STD metric were computed by averaging over the entire the 2D map. The computational cost was evaluated on a workstation with an NVIDIA GeForce RTX 3090Ti GPU (NVIDIA, Santa Clara, CA) and AMD Ryzen 9 5950X CPU (AMD, Santa Clara, CA). In the likelihood update step, we used a custom-written, PyTorch-compatible, CUDA-accelerated, distance-driven projector and back projector. Both the time and memory consumption are evaluated on the parallel sampling of 16 sets of projections. 

\subsubsection{Simulation Study Setup}
The proposed DPS CT reconstruction was first evaluated in a simulation study, emulating a system configuration as follows: The source-detector-distance (SDD) and source-axis-distance (SAD) were $1000$mm and $500$mm, respectively. An equal-spacing detector with $1024$ pixels and $1.0$mm pixel size was simulated. System blur $\textbf{B}$ was modeled as a shift-invariant Gaussian kernel with $\sigma=0.5$ pixel. For each 2D slice, CT measurements were simulated based on the nonlinear forward model in two scenarios: 1) low-mAs, and 2) sparse-view measurements. The low-mAs case used 720 projections with  $I_0 = 5\times10^{3}$ incident photons/pixel, and the sparse-view case used 72 projections with  $I_0 = 10^{5}$ incident photons/pixel. The ordered subset scheme was not used in the sparse-view reconstruction. The jumpstart reverse sampling was initialized using a FBP reconstruction of the raw projections. 

\begin{figure}[h]
\centering
\includegraphics[width=\linewidth]{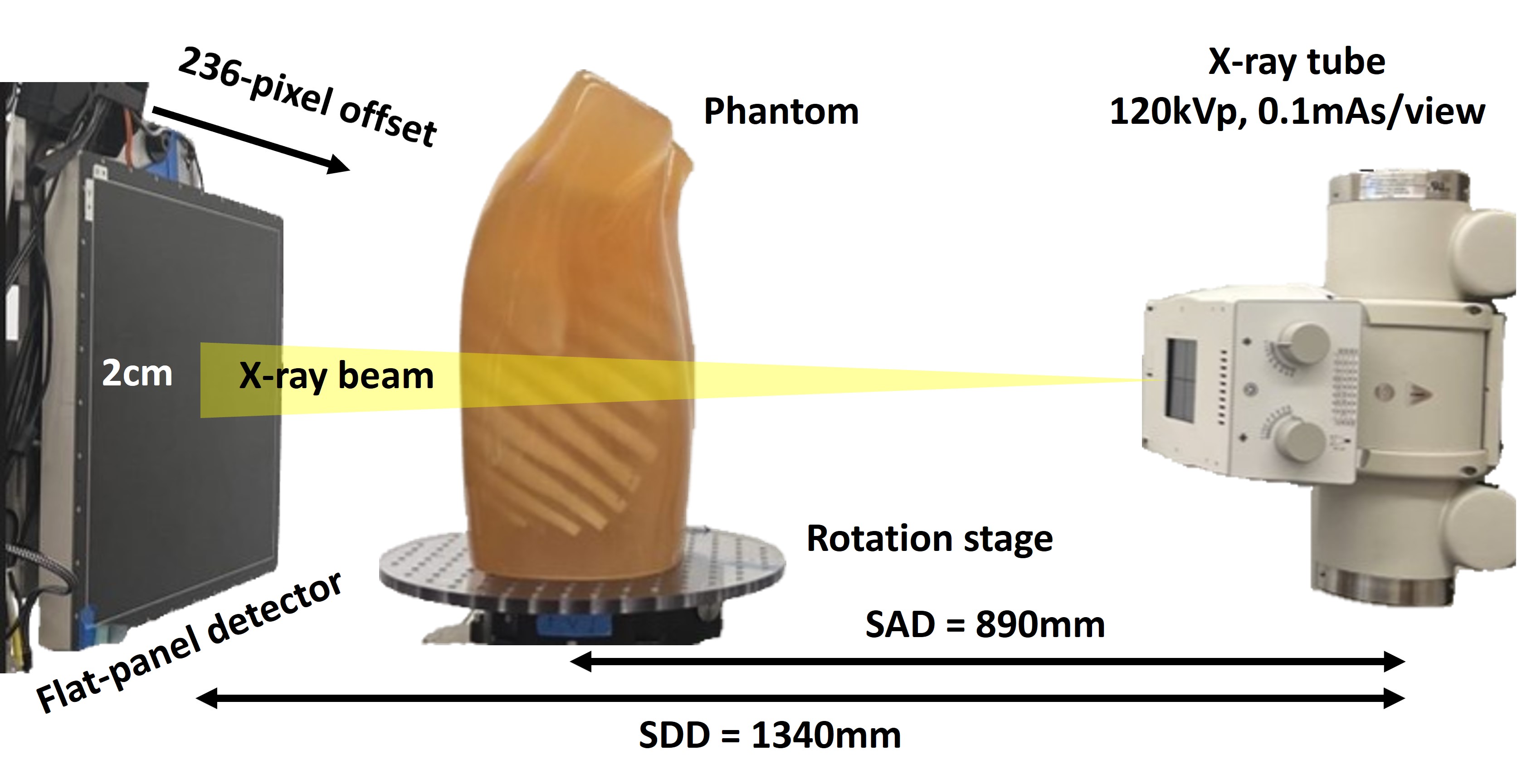}
\caption{The benchtop cone-beam CT system used in the phantom study.}
\label{fig:hardware} 
\end{figure}

\subsubsection{Phantom Study Setup}
The proposed algorithm is further validated on an anthropomorphic lung phantom (PH-1, Kyoto Kagaku, Japan). The projections were acquired on a benchtop cone-beam CT (CBCT) system (Fig.\ref{fig:hardware}) equipped with an x-ray tube (Rad-94, Varex Imaging, San Jose, CA) and a flat-panel detector (4343CB, Varex Imaging, San Jose, CA). The SDD and SAD were set to $890$mm and $1340$mm, respectively. The tube was operated in pulsed mode at $120$kVp and $0.1$mAs/view. The detector had a pixel size of $0.278$mm, and was positioned with a 236-pixel lateral offset to cover the entire phantom. The x-ray beam was vertically collimated to $2$cm width on the detector to minimize scatter, and only the central-slice measurements utilized for reconstruction. The normal-dose full-view scan acquired $720$ projections over $360^\circ$. We investigated the reconstruction performance with different dose levels ($100\%,20\%,10\%,5\%$ dose) and different numbers of views ($100\%,50\%,20\%,10\%$ views). The ground truth image was generated by averaging the FBP reconstructions of 8 repeated normal-dose full-view scans. System blur was not modelled in this phantom study since the reconstructed image voxel size ($0.8$mm) was much larger than the detector pixel size. 

\section{Results}
\label{sec:Res}
\begin{figure*}[h]
\centering
\includegraphics[width=\textwidth]{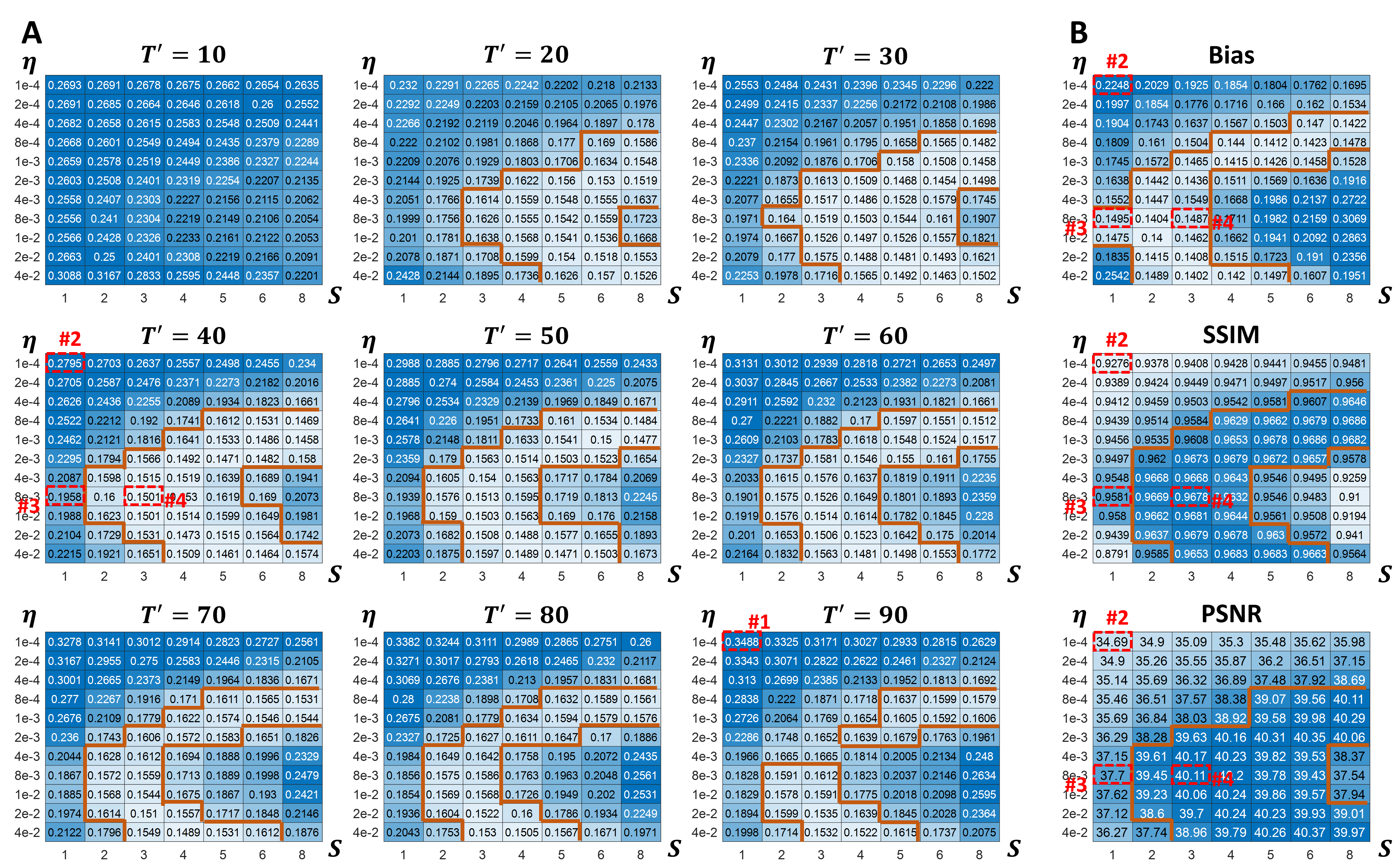}
\caption{Hyperparameters sweep results. (A) STD heatmap of proposed algorithm with different reconstruction hyperparameters (Unit: $10^{-3}mm^{-1}$). $T'$, $\eta$, $S$ represents jumpstart steps, likelihood update step size, and number of subset, respectively. The brown contours indicate the optimal regions for each metric, which are determined by the thresholds for each metric (STD$<0.1920$, Bias$<0.1600$, SSIM $>0.96$, PSNR $>39.00$). Four labelled reconstruction results indicated with red numbers and red dotted rectangles are displayed in Fig.\ref{fig:img_optim} for further investigation. (B) Bias, STD, SSIM, and PSNR for jumpstart step $T'=40$.}
\label{fig:std_optim} 
\end{figure*}

\begin{figure*}[h]
\centering
\includegraphics[width=\textwidth]{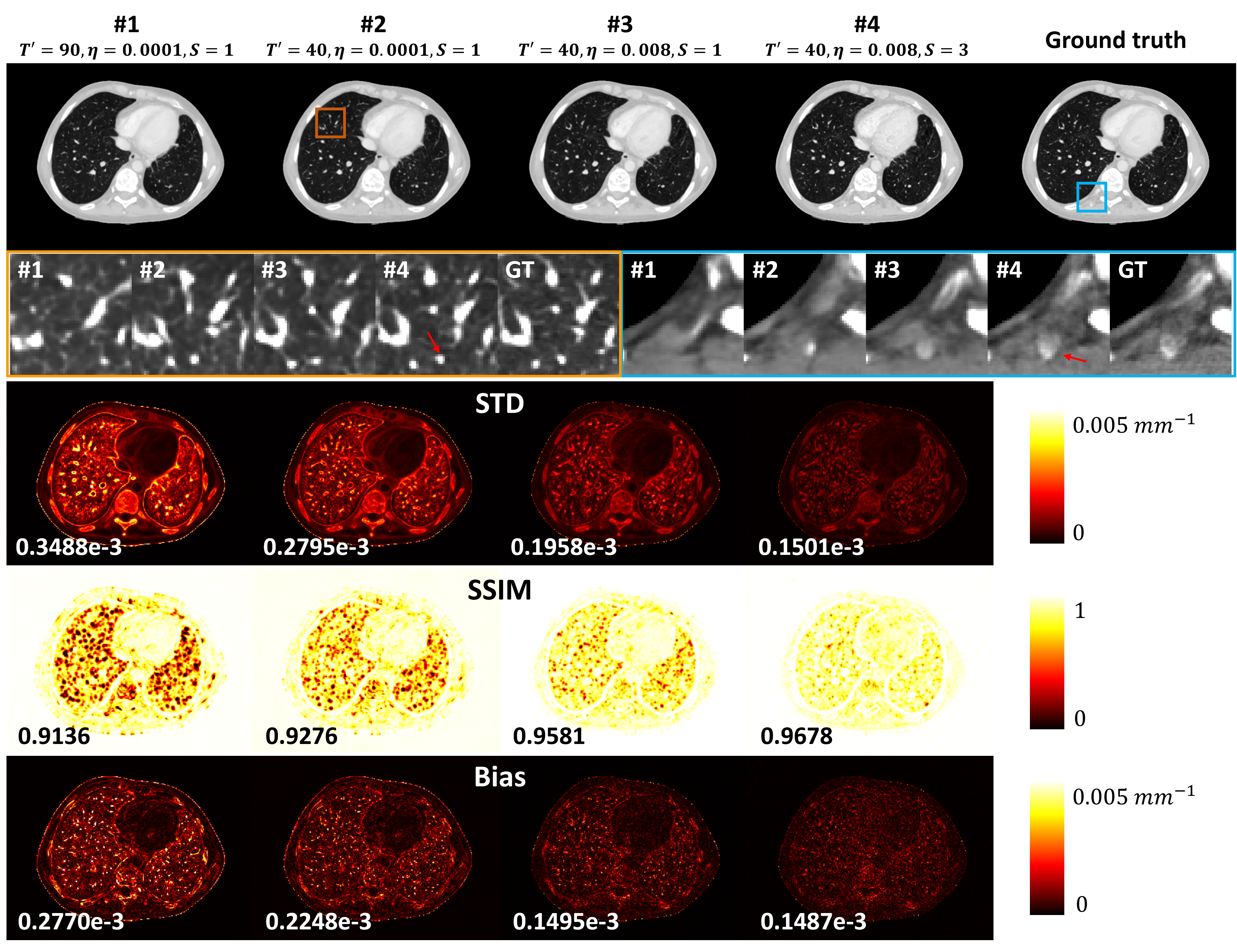}
\caption{Proposed low-mAs reconstruction results with hyperparameters labelled in Fig.\ref{fig:std_optim}. From \#1 to \#4: no parameter optimization, add jumpstart step optimization, add likelihood update step size optimization, add number of likelihood update optimization. Display window: full-size CT image: $[0,0.03]mm^{-1}$, lung ROI: $[0,0.01]mm^{-1}$, soft-tissue ROI: $[0.015,0.025]mm^{-1}$.}
\label{fig:img_optim}
\end{figure*}

\subsection{Hyperparameter Optimization}
Fig.\ref{fig:std_optim}A displays the hyperparameter sweep results for simulated low-mAs reconstructions using Algorithm 2. For this low-mAs scan setting, a $T'$ of around $30\sim50$ achieves minimal reconstruction variability, and the STD heatmap of these time steps displays similar trends for each $T'$. With only one likelihood update per time step, increasing the step size can encourage the data consistency thereby diminishing variability. However, there is a point where increased step size will lead to increased variability. The STD heatmap reveals that multi-step likelihood updates can further mitigate the variability. With increasing step size, the optimal number of likelihood updates first decreases, then increases, forming a similar C-shape region of low variability across a range of $T'$. 

Fig.\ref{fig:img_optim} further illustrates the impact of hyperparameter optimization on the proposed reconstruction. Beginning with unoptimized parameters ($T'=90, \eta=0.0001, S=1$), the STD map indicates large structural variability. Both the SSIM and bias maps reveal substantial reconstruction errors, which can be clearly visualized in the ROI images, particularly for tiny pulmonary vessels and low-contrast soft tissues. From \#1 to \#4, hyperparameter optimization suppresses reconstruction STD by $56.96\%$. This particular sequence optimizes \( T' \), then \( \eta \), then \( S \); and may be implemented as a succession of one-dimensional optimizations. Note this is close to the global optima of $0.1454$. Although the optimization aims to minimize the STD, it also improves the reconstruction accuracy, enhancing the depiction of the subtle details as indicated by the arrows. This improvement is also evidenced by the quantitative metrics summarized in Fig.\ref{fig:std_optim}B. All metrics show similar trends with different $S$ and $\eta$, suggesting that minimizing STD also benefits the reconstruction accuracy. We also note slight shifts in the 'optimal' regions for different metrics (e.g., bias is minimized with slightly lower $S$) suggesting that parameters can be further tuned based on user preference of the relative importance of different metrics. Bias, PSNR, and SSIM for other $T'$ are similarly related to STD and are not plotted for brevity.

\subsection{CT Reconstruction on Simulated Data}
\begin{figure*}[h]
\centering
\includegraphics[width=\textwidth]{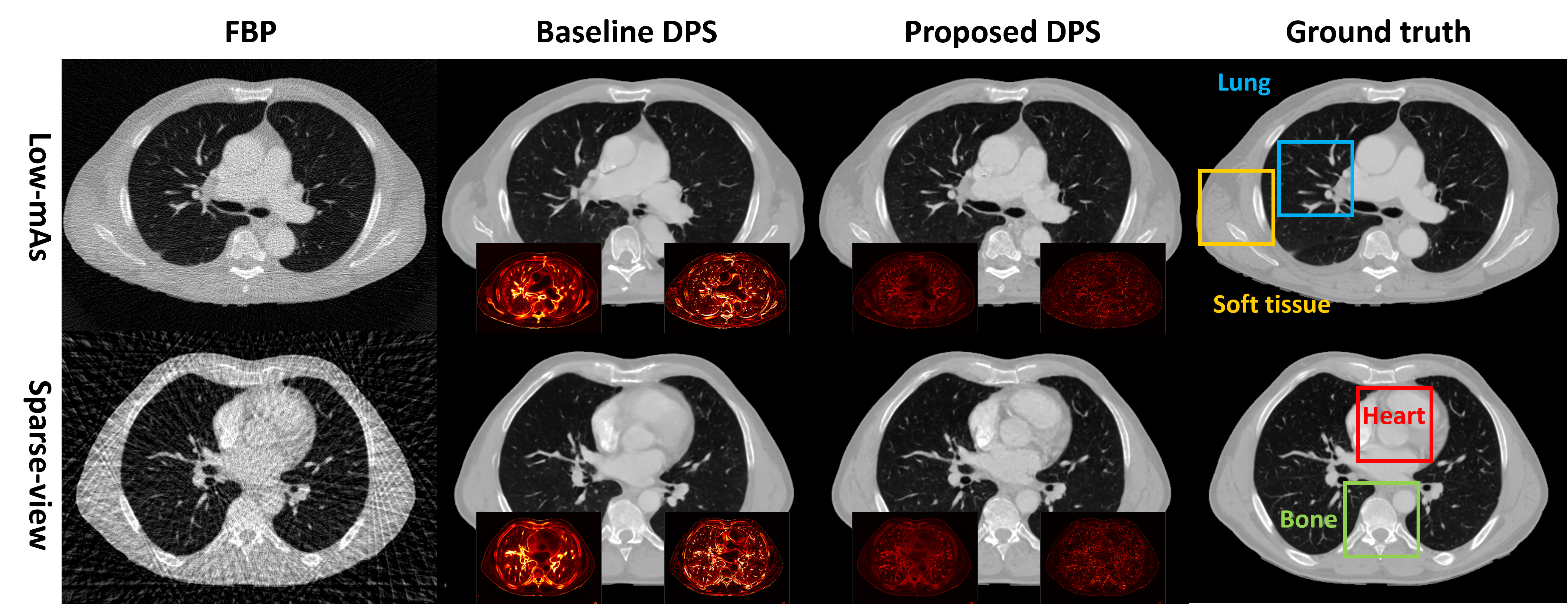}
\caption{Simulated low-mAs(top, $I_0 = 5\times10^3$) and sparse-view reconstruction(bottom, nView $= 72$). The STD and bias map are displayed on the bottom-left and bottom-right corner, respectively. Display window: CT image: $[0,0.03]mm^{-1}$, STD/bias map: $[0,0.005]mm^{-1}$}
\label{fig:simu_full} 
\end{figure*}

\begin{figure*}[h]
\centering
\includegraphics[width=\textwidth]{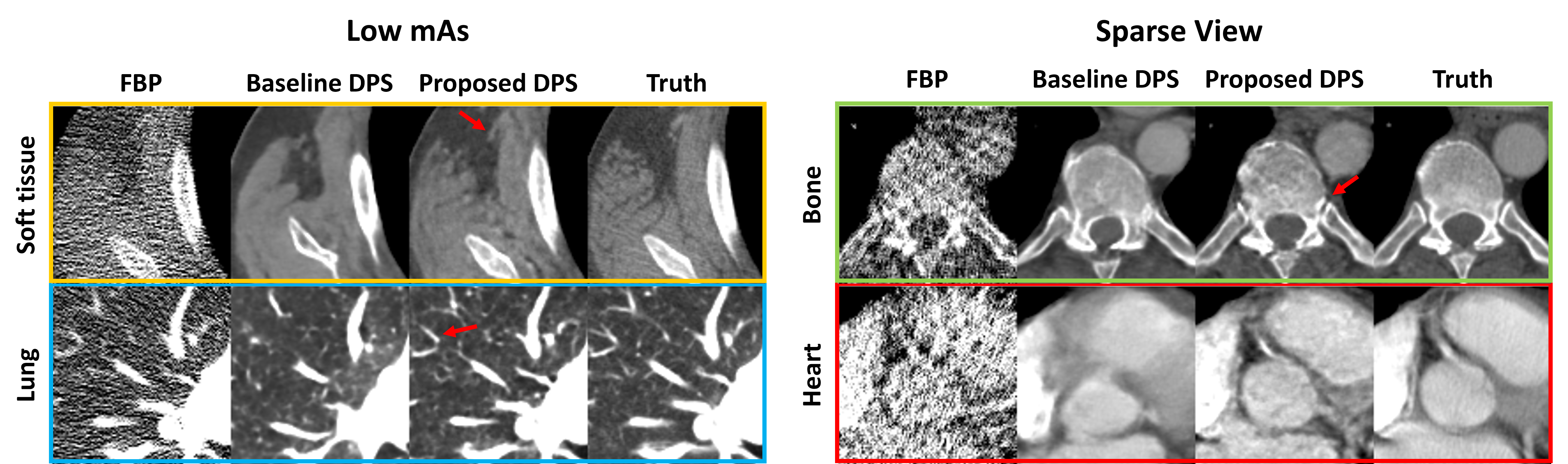}
\caption{ROI investigation of simulated CT reconstruction. Display window: soft-tissue/heart: $[0.015,0.025]mm^{-1}$, lung: $[0,0.01]mm^{-1}$, bone: $[0.015, 0.03]mm^{-1}$.}
\label{fig:simu_zoom} 
\end{figure*}
Simulated CT reconstruction results are summarized in Fig.\ref{fig:simu_full}. Both DPS and the proposed algorithm use parameters yielding minimal reconstruction STD. While baseline DPS is capable of generating realistic CT images for both low-mAs and sparse-view reconstruction, it suffers substantial structural bias in the heart and pulmonary vessel branches, as well as substantial anatomical variation around edges. In contrast, as seen in the bias and STD map, the proposed algorithm demonstrates less structural variation and improved reconstruction accuracy.   

Fig.\ref{fig:simu_zoom} displays four regions of interest (ROIs) - soft tissue, lung, bone, and heart - selected for detailed investigation, with corresponding quantitative metrics summarized in TABLE.\ref{tab:roi_metrics}. In the low-mAs setting, the FBP images are corrupted by streaky noise. While baseline DPS effectively suppresses image noise, it tends to misplace the tissue boundaries, as observed in the soft-tissue and lung ROIs. Compared with baseline DPS, the proposed algorithm achieves $26.03\%$ and $46.72\%$ higher PSNR and $13.72\%$ and $51.50\%$ higher SSIM on the soft tissue and lung ROIs, successfully recovering subtle muscle-fat boundaries and small pulmonary vessel branches as indicated by the arrows. 

The sparse-view scans introduce more severe streaking artifacts. Both baseline DPS and proposed DPS can mitigate the streaking artifacts. However, the proposed algorithm shows more accurate depiction of the spine shape and costotransverse joint over baseline DPS. In the heart ROI, the proposed algorithm delineates the boundary of the main pulmonary artery and the aorta which are blurred out by baseline DPS. More importantly, the proposed DPS accurately reconstructs the coronary vessel as well as the intravascular high-density segment, indicating superior reconstruction accuracy. This accuracy is also demonstrated by the quantitative metrics. The proposed DPS improves PSNR by $33.64\%$ and $23.68\%$ on the bone and heart ROIs and achieves a $52.20\%$ and $31.43\%$ STD reduction. 

\begin{table}[h]
\caption{Quantitative Metrics Evaluated on ROIs in Fig.\ref{fig:simu_zoom}}
\label{tab:roi_metrics}
\centering
\begin{tabular}{c c|c|c|c|c}
\hline
\multirow{2}{*}{} & \multirow{2}{*}{} & \multicolumn{2}{c|}{Low mAs} & \multicolumn{2}{c}{Sparse View} \\
\cline{3-6}
                            &                   & Soft tissue& Lung & Bone & Heart \\
\hline
\multirow{3}{*}{Baseline DPS}       & \multicolumn{1}{|c|}{STD}  & 0.63e-3  & 1.47e-3     &1.36e-3      &0.70e-3\\
                                    & \multicolumn{1}{|c|}{PSNR} & 28.4739    & 19.3431    & 23.5276    & 26.3828\\
                                    & \multicolumn{1}{|c|}{SSIM} & 0.7866     & 0.5285     & 0.6487     & 0.7224\\

\hline
\multirow{3}{*}{Propsoed DPS}       & \multicolumn{1}{|c|}{STD}  & 0.32e-3    & 0.60e-3    & 0.65e-3    & 0.48e-3\\
                                    & \multicolumn{1}{|c|}{PSNR} & 35.8879   & 28.3810   & 31.4429   & 32.6309\\
                                    & \multicolumn{1}{|c|}{SSIM} & 0.8945    & 0.8007    & 0.8446    & 0.8085\\
\hline
\end{tabular}
\end{table}

\subsection{CT Reconstruction on Real Data}
\begin{figure*}[h]
\centering
\includegraphics[width=\textwidth]{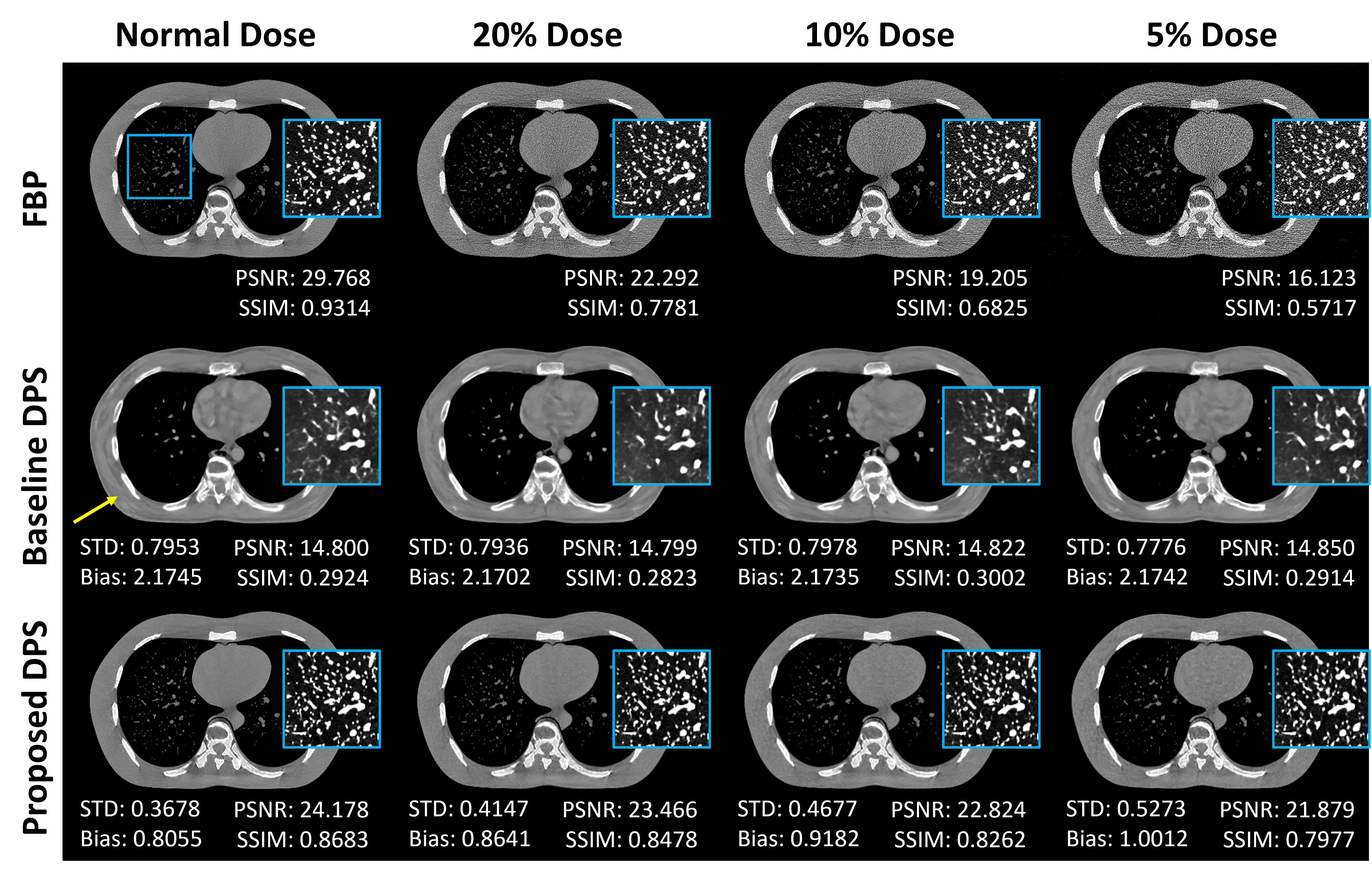}
\caption{Physical data reconstructions results with different dose levels. The STD (unit: $10^{-3}mm^{-1}$), bias (unit: $10^{-3}mm^{-1}$), PSNR, SSIM are computed on the lung ROI. Display window: Full size image: $[0.01,0.03]mm^{-1}$, lung ROI: $[0,0.01]mm^{-1}$.}
\label{fig:real_low} 
\end{figure*}

\begin{figure*}[h]
\centering
\includegraphics[width=\textwidth]{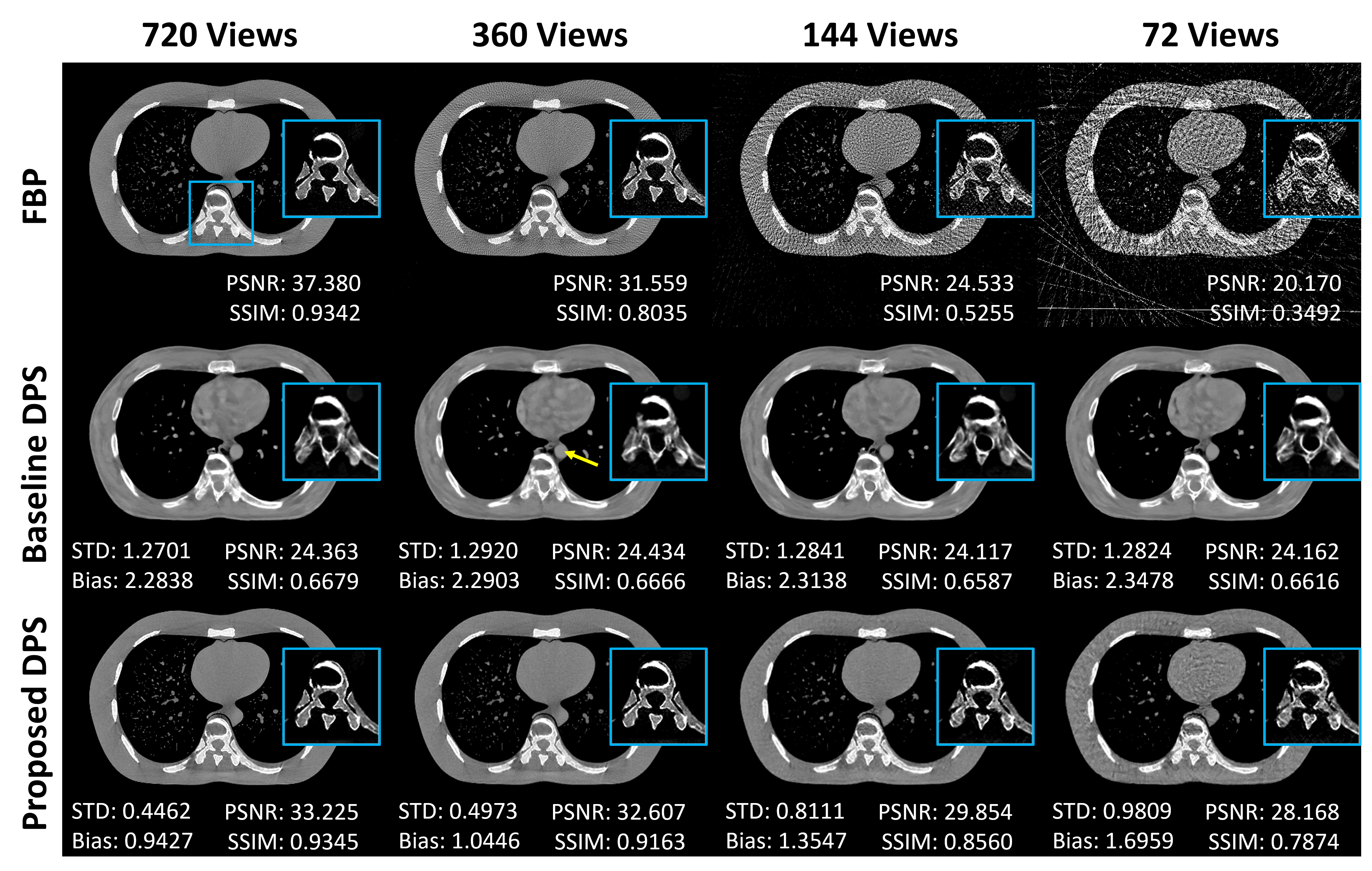}
\caption{Physical data reconstructions results with different number of views. The STD (unit: $10^{-3}mm^{-1}$), bias (unit: $10^{-3}mm^{-1}$), PSNR, SSIM are computed the spine ROI. Display window: Full size image: $[0.01,0.03]mm^{-1}$, spine ROI: $[0.020,0.035]mm^{-1}$.}
\label{fig:real_sparse} 
\end{figure*}

Fig.\ref{fig:real_low} and \ref{fig:real_sparse} display reconstructions of physical bench data. FBP results are added to visualize the different dose and down-sampling levels. Reconstruction on real data is a potentially more challenging test case since the anthropomorphic phantom is made of homogeneous materials, which does not strictly follow the training data distribution. As observed in Fig.\ref{fig:real_low}, DPS tends to generate structures inside the homogeneous heart and surrounding soft tissue. Interestingly, increasing the dose level does not necessarily improve the baseline DPS accuracy even with an optimized step size. This indicates that the baseline DPS highly relies on the diffusion prior and cannot fully use the physical measurements. In contrast, the proposed DPS algorithm modifies the likelihood update to enforce closer adherence to the physical measurement model to successfully reconstruct most of the vessel trees. Although the quality of the proposed reconstruction degrades with lower dose, only $5.60\%$ and $4.84\%$ reductions of PSNR and SSIM were observed with $10\%$ radiation dose. Even a reduction of dose to $5\%$ only leads to $<10\%$ lower SSIM/PSNR in the lung ROI, indicating the superior generalization ability of the proposed algorithm.

For reconstructions with different numbers of views, the baseline DPS also tends to create internal structures of the heart and misinterpret the soft tissue as descending aorta as pointed by the yellow arrow. Moreover, the depiction of the spine more closely resembles patient anatomy rather than the actual homogeneous phantom structures. We observed that the improved DPS algorithm achieves a more accurate spine depiction, enabling clear differentiation between cortical bone and cancellous bone. The proposed DPS surpassed FBP in terms of SSIM and PSNR in 360-view reconstructions. With a smaller number of views ($144/72$), the proposed method effectively mitigates both image noise and streaking artifacts, achieving $10\%$ view reconstructions while sacrificing only $15.22\%$ SSIM and $15.74\%$ PSNR. It is worth noting that both baseline DPS and the proposed DPS algorithm exhibit slight blurring in comparison to the FBP results. This discrepancy is attributed to the diffusion prior being trained on diagnostic CT images, which typically possess a lower resolution than the FPD-based CBCT.

\subsection{Computational Cost}
The computational cost of parallel 16-slice reconstructions is evaluated on the simulated low-mAs and sparse-view systems, with results summarized in TABLE.\ref{tab:cost}. Hyperparameters used for low-mAs reconstruction were $T' = 40, \eta = 0.008, S = 3$, and hyperparameters used for sparse-view reconstruction were $T' = 100, \eta = 0.02, S = 2$. The proposed DPS achieved a reconstruction speed of $0.99s/$slice and $1.48s/$slice on low-mAs and sparse-view reconstruction, respectively. Specifically, the jumpstart strategy skips over $\geq900$ starting time steps, shortening the time for diffusion sampling by $95.86\%$ and $89.94\%$ for low-mAs and sparse-view reconstruction, respectively. The time spent on likelihood updates also exhibited a remarkable speed-up of $97.48\%$ and $97.91\%$. We note that for the two scenarios, since the Jacobian approximation bypasses the network Jacobian computation, the proposed DPS saves $\sim40\%$ on GPU memory compared with baseline DPS. 

\begin{table}[h]
\caption{Computational cost of different reconstruction algorithms}
\label{tab:cost}
\centering
\begin{tabular}{c|c|c|c}
\hline
\multicolumn{2}{c|}{}  &  Baseline DPS & Proposed DPS \\
\hline
\multirow{4}{*}{Low mAs}            & Diffusion         & 188.2s     &  7.8s     \\
                                    & Likelihood update & 354.6s      & 8.1s     \\
                                    & Total time & 442.8s      & 15.9s     \\
                                    & Memory      & 17.5GB      & 10.5GB      \\

\hline
\multirow{4}{*}{Sparse view}        & Diffusion        & 187.8s     & 18.9s     \\
                                    & Likelihood update & 190.5s      & 4.8s      \\
                                    & Total time & 378.3s      & 23.7s     \\
                                    & Memory      & 17.4GB      & 10.3GB      \\
\hline
\end{tabular}
\end{table}

\section{Conclusion and Discussion}
\label{sec:Dis}
The DPS methodology has introduced a novel framework for nonlinear CT reconstruction by integrating a diffusion prior and an analytic system physical model. This innovative method shows promising performance with its ability to capture rich and generalized information about CT images to enhance reconstruction accuracy and retain realistic images. However, baseline DPS struggles with issues of large variability, hallucinations, and slow reconstruction speed. This work introduced a number of strategies designed to enhance the stability and efficiency of DPS CT reconstruction. Both simulation and phantom studies demonstrate that the proposed method can significantly reduce reconstruction variability and computational costs, greatly enhancing the practicality of DPS CT reconstruction.

SGM-based algorithms can suffer long sampling time due to the extensive number of reverse steps. This work introduces the jumpstart strategy for acceleration, which maintains the original DDPM reverse solver but skips most of the starting steps. Our experiments show that the jumpstart strategy reduces the required time steps to as few as $40$, substantially lowering time consumption and minimizing stochasticity during the reverse sampling. We note that Chung \textit{et al.}\cite{chung2022come} suggested a similar concept, however, the investigation was confined to linear problem, and depended on a pre-trained neural network for initialization. Our work extends this idea to the nonlinear scenario, and demonstrates even a coarse FBP initialization severely corrupted by noise and streaking artifacts is enough to boost the performance. It is reasonable to expect that the sampling step can be further reduced by integrating well-established pre- or post- processing techniques\cite{li2004nonlinear,galigekere1999techniques, kyriakou2010empirical, meyer2010normalized} to improve the initialization quality. Fast solvers like DDIM\cite{song2020denoising} and DPM-solver\cite{lu2022dpm} have been developed to decrease sampling steps while maintain the sampling quality. Future work will also investigate combining fast solvers with our improved strategies to furtherimprove the reconstruction speed and quality.

Variability is usually unfavorable for medical imaging. The jumpstart strategy mitigates variability by reducing the stochasicity during the reconstruction, while the modified likelihood update aims to enhance data consistency to restrict the variability. Recent studies\cite{chung2023fast,song2023solving,xia2023diffusion} have adopted a similar likelihood update approach by substituting gradient descent with an optimization problem that is approximately solved in a fixed number of iterations with classic optimization algorithm. This work did a exhaustive parameter sweep, illustrating that good data consistency can be achieved with fewer than $4$ iterations per time step, which significantly accelerates the likelihood update process, and the additional computational time may be effectively compensated by the ordered subset strategy. Future work can explore more MBIR strategies\cite{erdogan1999ordered,tilley2017penalized,tivnan2020preconditioned} to further accelerate and stabilize the DPS reconstruction.

Although a highly accurate description of the CT imaging process involves a nonlinear forward model, much research opts for a simplification to a linear model based on first-order Taylor expansions\cite{thibault2007three,wang2006penalized}. While the linearized model may be acceptable in my situations, we have observed performance degradation in DPS of very low exposures \cite{li2024ctreconstructionusingdiffusion}. Moreover, nonlinear models become much more important for extensions to more sophisticated CT models that include system blur\cite{tilley2017penalized}, spectral sensitivity\cite{tilley2019model}, scatter radiation, etc. The foundations laid in this work are expected to enable application in other imaging systems and scenarios where noise and sparsity would ordinarily prohibit good image estimates. This includes applications like spectral CT which seeks to jointly reconstruct and decompose images into material basis maps. Preliminary studies have already demonstrated the value of the proposed DPS methods on spectral material decomposition\cite{jiang2024ct}. We expect other modalities and data acquisition protocols to similarly benefit.

\appendices
\section{Score Network Jacobian Evaluation}
\label{sec:append_jacob}

\begin{figure*}[h]
\centering
\includegraphics[width=\textwidth]{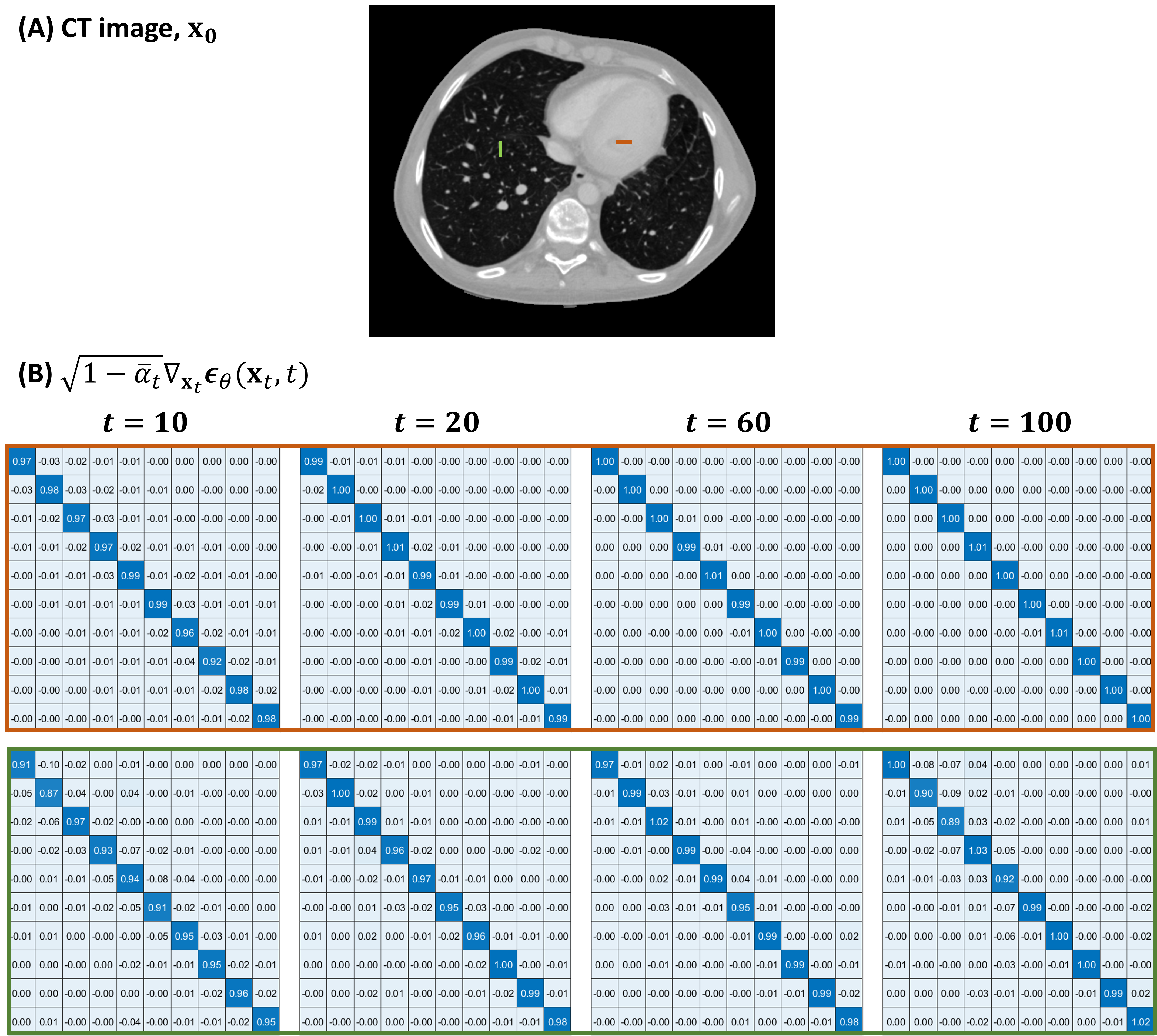}
\caption{A: CT image on which we evaluate the network Jacobian. B: the local network Jacobian evaluated on the segments in (A) at different time steps, each pixel of subplot represents the partial derivative of single pixel in network output with respect to the single pixel in the input. The Jacobian value is scaled by $\sqrt{1-\bar{\alpha}_t}$ to better illustrate $\nabla_{\textbf{x}_t}\boldsymbol{\epsilon}_\theta(\textbf{x}_t,t) \approx \textbf{I}/\sqrt{1-\bar{\alpha}_t}$.}
\label{fig:jacob} 
\end{figure*}

Fig \ref{fig:jacob} displays an example of a score network Jacobian $\nabla_{\textbf{x}_t}\boldsymbol{\epsilon}_\theta(\textbf{x}_t,t)$ for different time steps. Since the full Jacobian is very large, we only numerically evaluated the Jacobian on two segments indicated in Fig.\ref{fig:jacob}A, with results listed in Fig.\ref{fig:jacob}B. We observe that for $t$ from $10$ to $100$, $\sqrt{1-\bar{\alpha}_t}\nabla_{\textbf{x}_t}\boldsymbol{\epsilon}_\theta(\textbf{x}_t,t)$ can be well approximated by an identity matrix for both segments, which means $\nabla_{\textbf{x}_t}\boldsymbol{\epsilon}_\theta(\textbf{x}_t,t) \approx \textbf{I}/\sqrt{1-\bar{\alpha}_t}$. While not shown here, we have similar results for other positions.

\section*{Acknowledgment}
This work is supported by NIH grant R01CA249538.

\bibliography{report} 
\bibliographystyle{IEEEref}

\end{document}